\documentstyle[prd,aps,epsf,epsfig]{revtex} 
\begin{document}
\draft
\twocolumn[\hsize\textwidth\columnwidth\hsize\csname@twocolumnfalse\endcsname
%
\title{Measuring preferential attachment for evolving networks}

\author{H. Jeong, Z. N\'eda$^*$ and A.L. Barab\'asi}
\address{University of Notre Dame, Department of Physics, Notre Dame, IN 46616,
USA}

\maketitle
\centerline{\small (Last revised \today)}

\begin{abstract}
A key ingredient of current models proposed to capture
the topological evolution of complex networks is the hypothesis that 
highly connected nodes increase their connectivity faster than their 
less connected peers, a phenomenon
called preferential attachment. Measurements on four 
networks, namely the science citation network, Internet, actor collaboration
and science coauthorship network indicate that the
rate at which nodes acquire links depends on the node's
degree, offering direct quantitative support for the 
presence of preferential attachment.
We find that for the first two systems the attachment
rate depends linearly on the node degree, while for the latter two
the dependence follows a sublinear power law.  
\end{abstract}

\pacs{PACS numbers:89.65.-s, 89.75.-k, 05.10.-a }
\vspace{2pc}
]

\vspace{1cm}



Modeling the highly interconnected nature of various social, biological 
and communication systems as complex networks or graphs has
attracted much attention in the last few years. 
\cite{watts2,watts1,nature,physicaA,cohen,mendes,slanina,redner1,kertesz,physicaA2,PRE,newman1,vasquez,mendes2}. 
As for a long time these networks were 
modeled as completely random \cite{erdos,bollobas}, the recent interest is
motivated by the increasing
evidence that real network display short length-scale clustering
\cite{watts2,watts1}
and obey unexpected scaling laws \cite{nature,physicaA},
interpreted as signatures of deviation from randomness.  
Current approaches, using the tools of statistical physics 
\cite{mendes,redner1,kertesz} search for universalities both 
in the topology of these webs and in the 
dynamics governing their evolution. These efforts resulted in a
class of models that view networks as
evolving dynamical systems, rather than static graphs. 
Most evolving network
models \cite{physicaA,mendes,kertesz} are based on 
two ingredients \cite{physicaA}:
{\em growth} and {\em preferential attachment}. 
The {\em growth} hypothesis
suggests that networks continuously expand through the 
addition of new nodes and links between the nodes, while the
{\em preferential attachment} hypothesis states that the rate
$\Pi(k)$ with which a node with $k$ links
acquires new links is a monotonically increasing
function of $k$.
While most versions of such evolving network models
assume that $\Pi(k)$ is linear in $k$
\cite{physicaA,mendes,kertesz}, recently several authors 
proposed that $\Pi(k)$ could follow a power-law \cite{redner1,physicaA2}.
Consequently, the time evolution of the degree $k_i$ 
of node $i$ can be obtained from the first order differential equation
\begin{equation}
\frac{dk_i}{dt}=m \Pi(k_i),
\end{equation}
where $m$ is a constant and $\Pi(k)$ has the form
\begin{equation}
\Pi(k_i)=\frac{k_i^{\alpha}}{\sum_{j} k_j^{\alpha}}=C(t) k_i^{\alpha},
\end{equation}
with $\alpha>0$ an unknown scaling exponent.
For $\alpha=1$ these models reduce to the
scale-free model \cite{physicaA}, for which the degree distribution $P(k)$,
giving the probability that a node has $k$ links, follows $P(k)\propto
k^{-\gamma}$ with $\gamma=3$.
As Krapvisky, Redner and Leyvraz have shown \cite{redner1}, for $\alpha<1$
the degree distribution follows a stretched
exponential, while for $\alpha>1$ a gelation-like phenomenon is
expected, where
a single site links to nearly all other nodes. On the other hand the
hypothesis (2) raises a series of fundamental questions, that
are not yet supported directly by experimental data.
First, is preferential
attachment indeed present in real networks? I.e. does indeed $\Pi(k)$
depend on $k$, or it is $k$ independent, as assumed both by the Erd\H{o}s-R\'enyi
\cite{erdos} or
Watts-Strogatz \cite{watts1} models?
Second, if $\Pi(k)$ does indeed depend on $k$, what is
its functional form? Is it linear, as assumed in \cite{physicaA}, or follows
a power-law as suggested in \cite{redner1,physicaA2}? Could $\Pi(k)$ follow some unknown and yet
unexplored functional form?                                                                   

Here we offer the first direct attempt to answer these questions in
quantitative terms by proposing a numerical method that allows us
to extract the functional form of $\Pi(k)$ directly from dynamical data
on real evolving networks. Our measurements indicate that $\Pi(k)$ follows a power
law for all investigated networks. For the Internet
and the citation network we find $\alpha=1$, while for the science
collaboration network and the Hollywood actor network the 
results indicate sublinear attachment, i.e. $\alpha<1$. 
These results offer the crucial missing link for modeling
the dynamics of complex evolving networks. 

{\it Methods:} To measure $\Pi(k)$ we use                    
computerized data on the dynamics of large networks.
Consider a network for which we know the order in which each node and link joins the
system. According to (1) and (2) the function $\Pi(k)$ gives the
rate at which an existing node with $k$ links acquires new links
as the network grows. To measure $\Pi(k)$ we
need to monitor to which old node new nodes link to, in function of the degree
of the old node.
However, there is an important problem with this simple
approach: the normalization constant, $C(t)$, depends on the 
time at which a
given node joins the system, creating unwanted biases in the measurements. To avoid such
bias we study the attachment of new nodes within a relatively short
time-frame. Consider all nodes existing in the
system at time $T_0$, called "$T_0$ {\em nodes}". Next select a group of
"{\it $T_1$ nodes}", added between
$[T_1, T_1+\Delta T]$, where $\Delta T << T_1$ and $T_1>T_0$.
When a $T_1$ node  joins the system we record the degree $k$ of the 
$T_0$ node to which the new node links to. The 
histogram providing the number of links acquired by the $T_0$ 
nodes with exactly $k$ 
degree, after normalization, gives
the $\Pi(k,T_0, T_1)$ function.
If the growing network develops a stationary state,
$\Pi(k, T_0, T_1)$ should be independent of $T_0$ and $T_1$, and should 
depend on $k$ only, providing us the $\Pi(k)$ preferential attachment function.
As we are forced to use relatively short $\Delta T$ intervals, even for large 
networks with hundreds of thousands of nodes
$\Pi(k)$ has significant fluctuations, particularly for
large $k$.
To reduce the noise level, 
instead of $\Pi(k)$ we study the cumulative function:
\begin{equation}
\kappa(k)=\int_0^k \Pi(k) dk
.
\label{kappa}
\end{equation}
If $\Pi(k)$ follows (2), we expect that
\begin{equation}
\kappa(k)\propto k^{\alpha+1}.
\end{equation}

{\it Measurements:} The method described above can be applied to systems for which
the order in which the nodes are added to the network is known. 
In this respect we had access to four different computerized networks,
whose main parameters are shown in Table 1.

(1) In the coauthorship networks of 
neuro-science (NS) the nodes are scientists, two nodes 
being linked if they coauthored a paper together \cite{PRE}.
The database considered by us contains paper titles and authors
of all relevant journals in the field of NS published 
between 1991-98. Similar to other collaboration networks \cite{newman1}
the distribution $P(k)$ for NS follows a power-law.
Papers
published between 1991-9x are used to reveal the network topology up to
the considered 199x year, so that
papers published in year 199x+1 are used to measure $\Pi(k)$. 

(2) In the citation network the nodes are papers published in 1988 
in Physical Review Letters, and links represent the
citations these articles received.
$T_0$ is chosen as the year 1989.

(3) In the actor network nodes are actors which are linked if they
played together in a movie. The network investigated by us contains all movies and actors
from 1892 up to 1999 \cite{watts1,imd}. 
We determined $\Pi(k)$ for actors that debuted
between 1920 and 1940, i.e. $T_0$=1940. We followed the
evolution of the new links yearly
between 1942 and 1993.

(4) For the Internet data we investigated the nodes represent
Autonomous Systems (AS) and links are direct connections between them 
\cite{AS}. The available data follows the network evolution
from 1997 up to the present days. The function $\Pi(k)$ was determined
for the nodes existing in year 2000.

{\it Results:} The $\kappa(k)$ functions obtained for the discussed 
databases are shown in Figs.~1 and 2. If preferential attachment
is absent, i.e. $\Pi(k)$
is independent of $k$, we expect $\kappa(k)\propto k$.
In Figs.~1 and 2 we show as continuous line the linear fit. In each
of the investigated examples the increase of 
$\kappa(k)$ is faster
than linear,
offering direct evidence that preferential attachment is present in
each of the considered systems.
Furthermore, we
find that the curves follow a straight line on a log-log plot,
indicating that with a good approximation the power law hypothesis
(2) is valid. Note that apart from statistical fluctuations, 
the functional form of $\Pi(k)$ is independent of $T_0$, supporting the
stationary nature of the attachment process. There is some degree of variation,
however, when it comes to the value of the exponent $\alpha$.

\begin{figure}[h]
\epsfig{figure=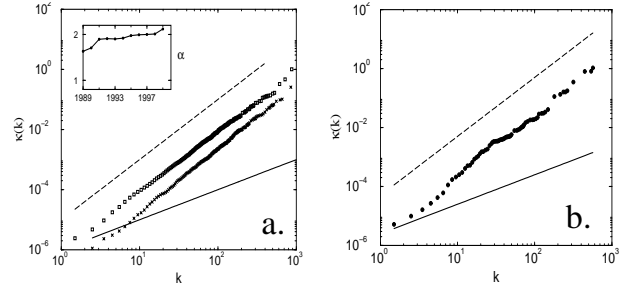,width=3.2in,angle=-0}
\caption{The $\kappa(k)$ function determined numerically for
the citation network (a)
and the Internet (b). In (a) the symbols from top to bottom
correspond to measurements made at $T_1=1991$ and $1995$, respectively.
For each curve we used $T_0=T_1-1$.
In the inset, we show the measured $\alpha$
exponents for each studied year which was obtained by fitting the whole
$\kappa(k)$ curve. 
For the
Internet (b) $\kappa(k)$ was determined using $T_0=1999$ and
$T_1=2000$, yielding $\alpha=1.05$ best exponent. 
In all measurements $\Delta T=1$ year.}
\label{fig1}
\end{figure}
\begin{figure}[h]
\epsfig{figure=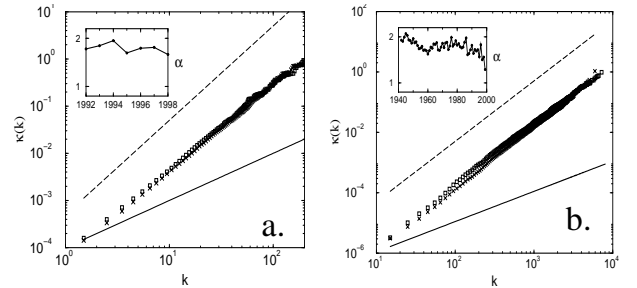,width=3.2in,angle=-0}
\caption{The $\kappa(k)$ function determined numerically for
the NS scientific collaboration (a)
and actor network (b). In (a) the symbols from top to bottom
correspond to measurements made at $T_1=1996$ and $1998$, respectively.
We have used $T_0=T_1-1$. 
In (b) the symbols from top to bottom correspond to
measurements made at $T_1=1950$ and $1960$, respectively. We
used as $T_0$ nodes, the actors present between 1920 to 1940.
In the inset we plotted the measured $\alpha$
exponents for each studied year. 
In all measurements $\Delta T=1$ year.}
\label{fig2}
\end{figure}

On Fig.~1 we present two $\kappa(k)$ curves  for the
citation network and for the Internet.  
For both networks we find that the slope of $\kappa(k)$
is very
close to two, shown as dashed line on the figure. For the Internet
where the measurement was performed for only one year, we obtain 
$\alpha=1.05$, while for the citation network we determined
$\kappa(k)$ for eight different years, obtaining the set of
$\alpha$ values shown in the Inset, indicating
$<\alpha>=0.95 \pm 0.1$.
Thus we conclude that for these two networks the linear ($\alpha=1$) 
preferential
attachment hypothesis offers a good approximation.

On the other hand, for the scientific collaboration and actor networks, we
find $\alpha<1$ (Fig.~2)
The set of $\alpha$ values for these
networks are summarized in the insets of Fig.~2. 
On average we get
$<\alpha>=0.81 \pm 0.1$ for the actor network, and $<\alpha>= 0.79 \pm 0.1$
for the scientific collaboration networks.

{\it Internal and external links:}
The observed sublinear behavior predicts that $P(k)$ for the 
systems shown in Fig.~2 should follow a stretched exponential \cite{redner1}. Nevertheless,
the measured $P(k)$ indicate that a power
law offers a
better fit. How can we than reconcile the nonlinear form of $\Pi(k)$ 
with  the measured $P(k)$? 
A potential resolution of this conflict lies in the
presence of internal links. 
For the scientific coauthorship network or the actor
web links appear not only from new nodes added
to the network, but as a result of new links connecting previously
existing nodes as well.
The method presented above allows us to investigate separately the attachment 
mechanisms of these 
distinct type of links.
For is, when determining $\Pi(k)$ we first limit the measurements only to
{\em external links}, i.e. links that
have been added to the system as a result of the appearance of new
nodes. Second, we focus only on new {\em internal links}, i.e. new links
that connect two previously  present
but disconnected nodes. For example, such internal link appears
when two researchers who have not published jointly  before, 
coauthor their first paper together or two actors, who did not act together
before, are joined in a new movie. In general, preferential attachment implies
that the probability that a new internal link appears between two nodes with
$k_1$ and $k_2$ degree
scales with the $k_1 k_2$ product \cite{PRE}. Focusing on the
actor network 
we find that both external and internal links follow preferential
attachment. However, the exponent $\alpha$ differs for the two
type of links. 
\begin{figure}[h]
\epsfig{figure=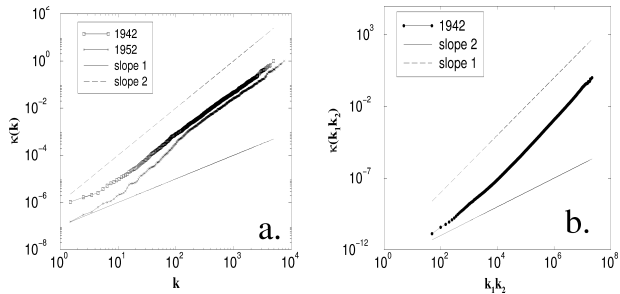,width=3.2in,angle=-0}
\caption{Preferential attachment of new nodes (a) and new internal
links (b) in the actor network. In (b) we plot the $\kappa(k_1k_2)$ 
function. Scaling of $\kappa(k)$
is illustrated for a few selected years. We obtain $0.7<\alpha<1$ for
the PA of new nodes (a) and $\alpha$ very close to one for the
internal links in the year chosen in figure (b).}
\label{fig3}
\end{figure}

From Fig.~3a we conclude that new incoming
nodes tend to link to the already existing nodes, following 
the functional form (2) with $\alpha<1$.
The $\kappa(k_1 k_2)$ function determined from the internal links
also follows 
a power-law. The  scaling is clearly not sublinear, and the exponent determined
from the assymptotic behavior is  close to two (Fig.~3b).
Thus the results indicate that the placement of the internal links is also governed
by preferential attachment, which scales linearly with $k$.
Similar results (not shown) have been obtained for the 
scientific collaboration web in neuroscience and mathematics \cite{PRE}.
Note that for the science citation network internal links are not allowed, and
the data resolution for the Internet does not allows to perform the same
analysis at this point.
We find that while the preferential attachment of the external
links is clearly sublinear, the internal links follow a close to linear
behavior. As in both the actor network and the scientific collaboration
network the number of internal links far out-weight the external links, we
believe that in the asymptotic limit the internal attachment is the
one that drives the shape of the $P(k)$ distribution, eventually being
responsible for its power-law form. These measurements raise several
interesting possibilities for the analytical treatment of the complex
coexistence of internal and external links, that could shed further
light on the evolution of complex networks.

{\it Initial attractiveness:}
Dorogovtsev, Mendes and Samukhin have recently suggested that
in order to account for the
fact that even nodes with no links can acquire links,
$\Pi(k)$ should have an additive term, $k_0$, called initial
attractiveness \cite{mendes} so that, $\Pi(k)\propto k_0+k^{\alpha}$.
For $\alpha=1$ it has been demonstrated the the degree exponent, $\gamma$,
depends continuously on $k_0$. In principle, having the functional form
of $\Pi(k)$ allows us to determine $k_0$ as well. We inspected the form of
$\Pi(k)$, which indeed does indicate that a nonzero $k_0$ is present. 
On the other hand the available
statistics was not sufficient to determine unambiguously the value of $k_0$.
In any case, our estimates indicate that $k_0$ is rather small, 
in the $10^{-6}$ range, thus its presence has no effect on the
scaling of $\kappa(k)$ at large $k$.
Nevertheless, the nonzero $k_0$ plays an important role in
starting the evolution of the node connectivity, since in its absence no
disconnected node could acquire initial links.

\vspace{0.25in}
{\small
\begin{tabular}{|c||c|c|c|}
\hline 
\multicolumn{1}{|c||}{Database} & \multicolumn{1}{|c|}{\# nodes} 
& \multicolumn{1}{|c|}{\# links} & \multicolumn{1}{|c|}{$\alpha$}
\\ \hline \hline 
Citation        &   1736         &  83252  &  $0.95 \pm 0.1$  \\ \hline 
Internet        &   12409        & 13445   &   $1.05$         \\ \hline
Collaboration   &   209293       & 3534724 &  $0.79\pm 0.1$   \\ \hline
Actor           &   392340       & 33646882&  $0.81\pm0.1$    \\ \hline 
\end{tabular}
}
\vspace{0.25in}

{\small Table 1. Summary of the investigated database, showing the number of nodes,
links, and the average value of the obtained exponent $\alpha$.}

\vspace{0.25in}

In summary, our measurement offer direct confirmation for the existence
of preferential attachment for rather different real evolving networks.
The emerging picture is more complex, however, than originally assumed
in \cite{physicaA}. We
find that for all the networks we studied Eq. (2) gives a good fit for 
$\Pi(k)$, implying that $\Pi(k)$ follows a power-law.
The exponent $\alpha$, however, is system
dependent: while for the Internet and the citation network a linear
$\Pi(k)$ offers a reasonable fit, for the actor network and collaboration web the
attachment rate is sublinear. These results give firmer foundation for the
evolving network models, that have been studied extensively to describe
the dynamics of real evolving networks.
But they also pose an important question: what is the
microscopic origin of preferential attachment? What determines the exponent $\alpha$ in general?
While some preliminary answers have been proposed \cite{vasquez,mendes2}, a good
understanding of this fundamental question is still lacking.

We acknowledge useful discussion with R. Albert, I. Der\'enyi,
and T. Vicsek. This research was supported by
NSF, PHY-9988674 and CAREER DMR97-01998.


\newpage


\begin{references}

\bibitem[*]{NZ}
{\em on leave from:} Babes-Bolyai University, Dept. of The\-o\-re\-ti\-cal Physics,
str. Kogalniceanu 1, RO-3400, Cluj-Napoca, Romania.

\bibitem{watts2} D.J. Watts, Small World (Princeton University Press,
Princeton, 1999); S. H. Strogatz, Nature {\bf 410}, 268 (2001); M.E.J. Newman, 
J. Stat. Phys. {\bf 101}, 819 (2000)
\bibitem{watts1} D. J. Watts and S.H. Strogatz, Nature {\bf 393}, 440 (1998)

\bibitem{nature} R. Albert, H. Jeong and A.-L. Barab\'asi, Nature {\bf 401},
130 (1999)

\bibitem{physicaA} A.-L. Barab\'asi and R. Albert, Science {\bf 286}, 509
(1999); A.-L. Barab\'asi, R. Albert and H. Jeong, Physica A
{\bf 272}, 173 (1999)

\bibitem{cohen} R. Cohen, K. Erez, D. ben-Avraham and S. Havlin, Phys. Rev.
Lett. {\bf 85}, 4626 (2000)
\bibitem{mendes} S. N. Dorogovtsev, J.F.F. Mendes and A. Samukhin, 
Phys. Rev. Lett. {\bf 85}, 4633 (2000) 

\bibitem{slanina} F. Slanina and M. Kotrla, Phys. Rev. E {\bf 62}, 6170
(2000)

\bibitem{redner1} P.L. Krapvisky, S. Redner and F. Leyvraz, 
Phys. Rev. Lett. {\bf 85},
4629 (2000)
\bibitem{kertesz} L. Kullmann and J. Kert\'esz, cond-mat/0012410

\bibitem{physicaA2} A.-L. Barab\'asi, R. Albert and H. Jeong, Physica A
{\bf 281}, 69 (2000)
  
\bibitem{PRE} A.-L. Barab\'asi, H. Jeong, Z. N\'eda, E. Ravasz, A. Schubert and T.
Vicsek, preprint 2001 
\bibitem{newman1} M.E.J. Newman Proc. Natl. Acad. Sci. USA, {\bf 98}, 404 (2001);
cond-mat/0007214
\bibitem{vasquez} A. Vazquez, cond-mat/0006132 
\bibitem{mendes2}  N. Dorogovtsev, J.F.F. Mendes and A. Samukhin,
cond-mat/0011115
\bibitem{erdos} P. Erd\H{o}s and A. R\'enyi, Publicationes
Mathematicae {\bf 6}, 290 (1959); P. Erd\H{o}s and A. R\'enyi, Acta Math. Sci.
Hung. {\bf 12}, 261 (1961)
\bibitem{bollobas} B. Bollob\'as, Random Graphs (Academic, London, 1985)

  
\bibitem{webofscience} http://www.webofscience.com

\bibitem{imd} http://www.imdb.com

\bibitem{AS} http://moat.nlanr.net/infrastructure.html

\end{references}
\end{document}